\documentclass[prd,twocolumn,amsmath,amssymb,superscriptaddress,floatfix,nofootinbib]{revtex4}
\usepackage{mathrsfs,bm}
\usepackage{longtable,lscape}
\usepackage{txfonts}
\usepackage{amssymb}
\usepackage{indentfirst}
\usepackage{graphicx,,booktabs}
\usepackage{multirow, overpic}
\usepackage{color}
\usepackage{amssymb}

\usepackage{color}
\usepackage[bookmarks=true,bookmarksopen=false,plainpages=false,breaklinks=true,
   bookmarksnumbered=true,hypertexnames=false,
   filecolor=blue,urlcolor=three,menucolor=three,
   linkcolor=three,citecolor=blueone, colorlinks,
   anchorcolor=blue,runcolor=pink,frenchlinks=red
   pdfstartview=FitH,pdftitle=title,%
   pdfauthor=author]{hyperref}
   
\allowdisplaybreaks[4]

\definecolor{cover}{rgb}{0.77,0.87,0.88}
\definecolor{blueone}{rgb}{0.1,0.1,.7}
\definecolor{citec}{rgb}{0.14,0.47,0.09}
\definecolor{two}{rgb}{0.0,0.5,0.}
\definecolor{three}{rgb}{.5,.1,0.15}

\begin{document}
\title{Induced fission-like process  of hadronic molecular states}

\author{Jun He}
\email{junhe@njnu.edu.cn}
\affiliation{Department of  Physics and Institute of Theoretical Physics, Nanjing Normal University,
Nanjing 210097, China}
\affiliation{
Lanzhou Center for Theoretical Physics, Lanzhou University, Lanzhou 730000, China
}

\author{Dian-Yong Chen}
\email{chendy@seu.edu.cn}
\affiliation{School of Physics, Southeast University, Nanjing 210094,  China}
\affiliation{
Lanzhou Center for Theoretical Physics, Lanzhou University, Lanzhou 730000, China
}

\author{Zhan-Wei Liu}
\email{liuzhanwei@lzu.edu.cn}
\affiliation{School of Physical Science and Technology, Lanzhou University, Lanzhou 730000, China}
\affiliation{
Lanzhou Center for Theoretical Physics, Lanzhou University, Lanzhou 730000, China
}

\author{Xiang Liu}
\email{xiangliu@lzu.edu.cn (Corresponding author)}
\affiliation{School of Physical Science and Technology, Lanzhou University, Lanzhou 730000, China}
\affiliation{
Lanzhou Center for Theoretical Physics, Lanzhou University, Lanzhou 730000, China
}

\date{\today} \begin{abstract} In this work, we predict a new physical
phenomenon, induced fission-like process  and chain reaction of hadronic molecular states. As a
molecular state, if induced by a $D$ meson, the $X(3872)$ can split into
$D\bar{D}$ final state which is forbidden due to the spin-parity conservation.
The breeding of the $D$ meson of the reaction, such as $D^0X(3872)\to
D^0\bar{D}^0D^0$, makes the chain reaction of $X(3872)$ matter possible.  We
estimate the cross section of the $D$ meson induced fission-like process of $X(3872)$ into
two $D$ mesons.  With very small $D^0$ beam momentum of 1~eV, the total cross
section reaches an order of 1000~b, and decreases rapidly with the increasing of
beam momentum.  With the transition of  $D^*$ meson in molecular states to a $D$
meson, the $X(3872)$ can release large energy, which is acquired by the final
mesons. The momentum distributions of the final $D$ mesons are analyzed. In the
laboratory frame, the spectator $D$ meson in molecular state concentrates in the
low momentum area. The energy from the transition frim $D^*$ to $D$ meson is
mainly acquired by two scattered $D$ mesons.   The results suggest that the $D$
meson environment will lead to the induced fission-like process and chain reaction of the
$X(3827)$. Such phenomenon can be extended to other hadronic molecular states.
\end{abstract}

\maketitle

\section{Introduction}

$X(3872)$ is the first observed hadronic molecule candidate with a mass very
close to the total mass of a ground state and an excited charm
meson~\cite{Belle:2003nnu}. Hence, it is suggested to be a hadronic molecular
state composed of two (anti)charm mesons~\cite{Tornqvist:2004qy}. Its wave
function can be written as~\cite{Liu:2008fh},
\begin{eqnarray}
|X\rangle&=&\frac{1}{\sqrt{2}}(|\bar{D}^{*0}{D}^0\rangle-|D^{*0}\bar{D}^0\rangle),\label{wf}
\end{eqnarray}
 $X(3872)$ carries the spin parity $J^P  = 1^+$, which cannot
decay into a $\bar{D}^0 D^0$ pair due to the conservation of spin parity. In
fact, if we do not consider the small decay probabilities of the $\bar{D}^{*0}$
and $D^0$ mesons, this hadronic molecular state should be stable, similar to a
deuteron.

In nuclear physics, if induced by a nucleon, a nucleus can split into two or
lighter nuclei by recombination of nucleons, even the spontaneous fission is forbidden~\cite{Meitner:1939gwm}.  If more
neutrons can be produced from the neutron-induced nuclear fission, that is, a
chain reaction happens. 
Hence, it is interesting to see if the $X(3872)$ can decay into a $\bar{D}^0
D^0$ pair if induced by an additional $D^0$ meson. 
Since the molecular state is composed of two hadrons, the
same type of fission of heavy nucleus is impossible for  molecular state.
However, the kinds of constituent hadrons of molecular states are much richer
than nucleus, which contains only nucleons~\cite{Chen:2020aos,Liu:2021pdu,Li:2021zbw}. Such property will lead to new type
of induced fission of molecular state compared with nucleus.

\begin{figure}[h!]
  \centering
  \includegraphics[bb=50 50 950 460,scale=.28,clip]{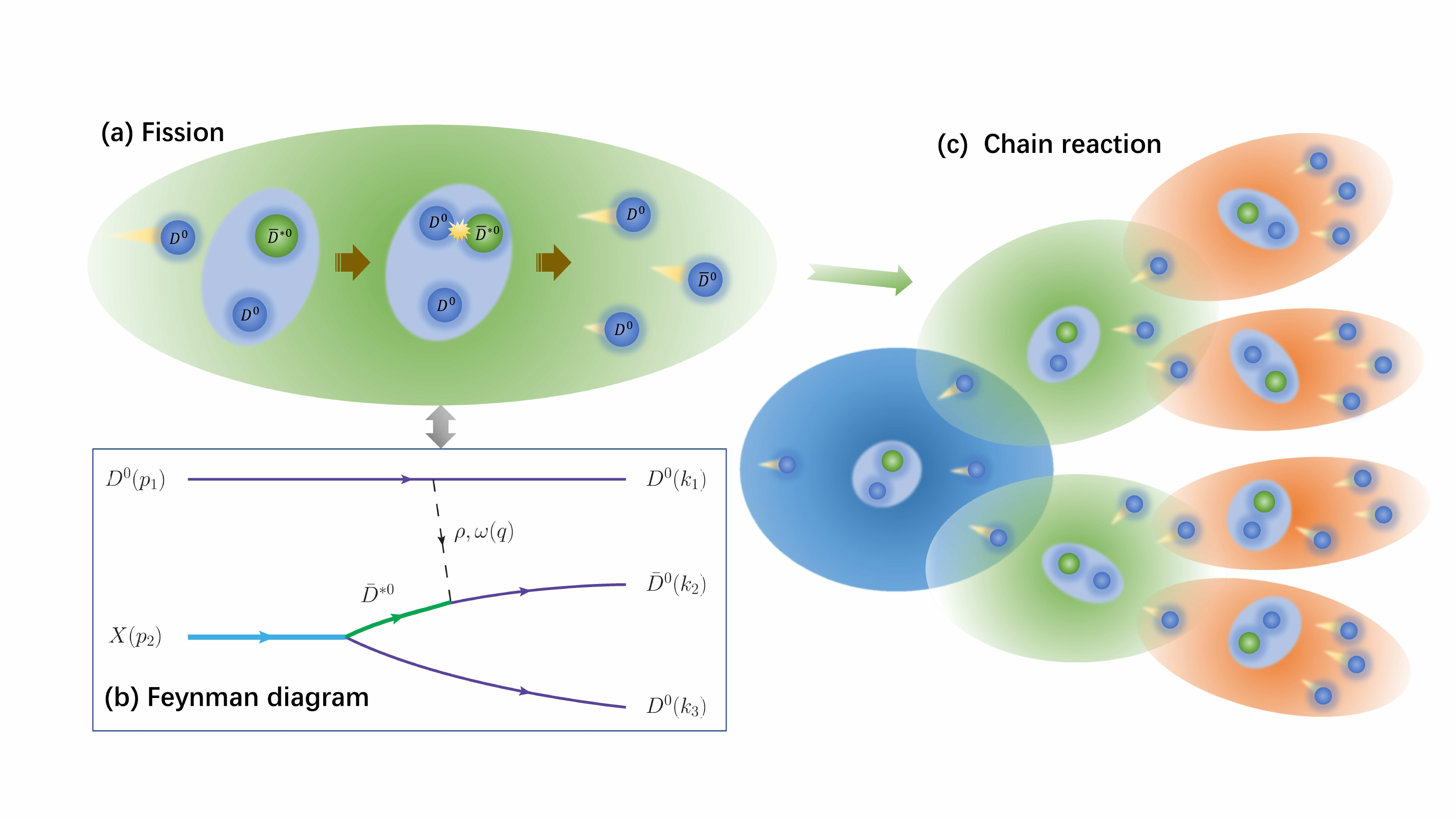}
  \caption{Schematic depiction of the induced fission-like process and chain reaction of $X (3872)$. (a) The induced fission-like process of $X (3872)$ with an incoming $D^0$ meson. The blue and green balls denote $D^0$ and $\bar{D}^{*0}$ mesons, respectively. The collision happens between the incoming $D^0$ meson and $\bar{D}^{*0}$ meson in the $X(3872)$ which is denoted by a small oval. (b) The Feynman diagram corresponding to the induced fission-like process in (a) with $\rho/\omega$ exchange. (c) The chain reaction of $X(3872)$ matter. The large oval means an induced fission-like process in (a). The overlap of the ovals and the meson therein means that the final meson of previous fission-like becomes the incoming meson of next fission-like process. }\label{picture}
  \end{figure}

In Ref.~\cite{He:2022rta}, we studied the nucleon-induced fission-like process of the
$T^+_{cc}$. With interaction of a nucleon, the  forbidden decay of $T^+_{cc}$ to
two $D$ mesons is allowed. Such phenomenon can be also expected in the case of
$X(3872)$. Furthermore, if we use the $D$ meson to attack the $X(3872)$, the
produced $D$ mesons can be taken as the inducing particle in the sequence
reaction, which is more analogous to the nuclear fission and chain reaction.
Hence, in the current work,  we will study decay of $X(3872)$ induced by an
additional $D^0$ meson with more $D$ mesons produced. The explicit scheme is
shown on the upper left of Fig.~\ref{picture}. $X(3872)$ can be either the
$\bar{D}^{*0} D^0$ molecule or the $D^{*0}\bar{D}^0$ molecule with equal
probability. That is, it is the superposition of the two molecular states that
can be understood and treated analogously, and thereafter, we take the first
term $\bar{D}^{*0} D^0$ as representative.

An incoming $D^0$ meson is used to bombard the molecular state $X(3872)$, which produces an intermediate state as $D^0 \bar{D}^{*0} D^0$ that can decay into three ground mesons as
\begin{align}
&D^0+X(3872)\to D^0\bar{D}^{*0}D^0\to 2{D}^0+\bar{D}^0.\label{X3872 fission}
\end{align}
In the reaction, the incoming $D^0$ meson is analogous to the neutron in nuclear fission. However, in the current reaction, the $D^{*0}$ meosn in $X(3872)$ transmuted to a $D^0$ meson, which is different from the nuclear fission where only recombination of constituents happens.  We define such a reaction as  the induced fission-like process of a molecular state.
The chain reaction is important to make nuclear fission proceed continuously. If the induced fission-like process occurs in $X(3872)$  as shown in the right part of Fig.~\ref{picture}, the produced $D^0$ mesons will induce the fission-like process of more molecular states. Thus, the chain reaction is also possible for hadronic molecular states.

\section{Cross sections} 

The probability of induced fission-like process can be
estimated by the cross section $\sigma$. In the induced fission-like process of
$X(3872)$, the incoming $D^0$ meson strikes on the $\bar{D}^{*0}$ meson in
$X(3872)$ while the interior $D^0$  meson is as a spectator, and then one
$\bar{D}^0$ and two $D^0$ mesons are produced. This can be described by a
Feynman diagram within field theory, as shown on the bottom left of
Fig.~\ref{picture}.  From the diagram, the $\bar{D}^{*0}$ meson in $X(3872)$
exchanges a $\rho/\omega$ meson with the incoming $D^0$ meson and transforms
into a $\bar{D}^0$ meson.  For meson-induced fission-like process of a molecular state it is
logical to use the rest frame of the molecular state, that is, the so-called
laboratory frame. In this reference frame, the cross section for the reaction
$D^0+X\to 2D^0+\bar{D}^0$ is expressed as,
\begin{align}
	d\sigma=\frac{1}{4[(p_1\cdot p_2)^2-m_1^2m_2^2]^{1/2}}\frac{1}{3}\sum_\lambda|{\cal M}_\lambda|^2d\Phi_3\frac{1}{2},\label{cross}
\end{align}
where the $p_{1,2}$ and $m_{1,2}$ are the momentum and mass of incoming $D^0$ meson or $X(3872)$. The phase space $d\Phi_3$ is produced with the help of GENEV code in FAWL as 
$R_3
	=(2\pi)^{5}d\Phi_3=\prod^3_i\frac{d^3k_i}{2E_i}\delta^4(\sum^n_ik_i-P)
$
where the $k_i$ and $E_i$ are the momentum and energy of final particle $i$. The mechanism of the fission-like process reaction can be described by an amplitude ${\cal M}_\lambda$ with $\lambda$ being the helicity of $X(3872)$. For the first term of the wave function in Eq. (\ref{wf}), the amplitude can be written as
\begin{align}
{\cal M}_\lambda=\frac{\sum_{\lambda_{\bar{D}^{*0}}}{\cal A}_{\lambda,\lambda_{\bar{D}^{*0}}}(X\to \bar{D}^{*0}{D^0}){\cal A}_{\lambda_{\bar{D}^{*0}}}(\bar{D}^{*0}{D}^0\to \bar{D}^0D^0)}{p^2-m^2_{D^*}},
\end{align}
where the different helicities for intermediate $\bar{D}^{*0}$ meson $\lambda_{\bar{D}^{*0}}$ should be summed up and the helictites will be omitted if not necessary. 

In the literature~\cite{Braaten:2004fk}, the $X(3872)$ splitting into $\bar{D}^{*0} D^0$ can be related to the scattering of $\bar{D}^{*0} D^0$. The coupling of the molecular state to its constituents can be related to  binding energy~\cite{Weinberg:1962hj}. Hence, the amplitude for $X(3872)$ splitting into $\bar{D}^{*0} D^0$ is determined by the scattering length $a$ as~\cite{Braaten:2004fk},
\begin{align}
{\cal A}_{\lambda,\lambda_{\bar{D}^{*0}}}(X\to \bar{D}^{*0}{D}^0)=	\sqrt{\frac{16\pi m_Xm_{D^*}m_D}{\mu^2a}}\epsilon_{X\lambda}\cdot\epsilon_{\lambda_{\bar{D}^{*0}}},
\end{align}
where $m_{X,D^*,D}$ is the mass of $X(3872)$, the constituent $ \bar{D}^{*0}$, or $D^0$. The $\epsilon_X$ and $\epsilon$ are the polarized vectors for $X(3872)$ and $ \bar{D}^{*0}$.  Scattering length $a=1/\sqrt{2\mu E_B}$ with the reduced mass $\mu=m_{D}m_{D^*}/(m_D+m_{D^*})$ and the $E_B$ being the binding energy. 

The propagator of $\bar{D}^{*0}$ in laboratory frame, where the $X(3872)$ is static, can be written as 
\begin{align}
\frac{1}{p^2-m^2_{D^*}}=\frac{1}{(m_X-E_3-E_{D^*})(m_X-E_3+E_{D^*})}.
\end{align}
Before  being struck, the $X(3872)$ is static and the binding energy is very small, which suggest that the momenta of the constituent mesons is small. The energy of two constituent mesons can be safely approximated as  $E_3=m_D^2+{\bm k}_3^2/2m_D$ and $E_{D^*}=m_{D^*}^2+{\bm k}_3^2/2m_{D^*}$. 
As in Ref.~\cite{He:2011ed}, the amplitudes  for $X(3872)$  splitting with the propagator of $D^*$ meson can be expressed with wave function of $X(3872)$ as 
\begin{align}
\frac{{\cal A}(X\to \bar{D}^{*0}{D}^0)}{p^2-m^2_{D^*}}&\simeq&-\frac{\sqrt{8m_Xm_{D^*}m_D}}{m_X-m_D+m_{D^*}}\psi({\bm k}_3)\epsilon_X\cdot\epsilon^*,
\end{align}
where wave function is\begin{align}
\psi({\bm k})=\sqrt{8\pi\over a}\frac{1}{{\bm k}^2+1/a^2},
\end{align} 
with normalization $\int d^3k/(2\pi)^3|\psi(k)|^2=1$.  Such wave function is consistent with the wave function adopted by Voloshin in coordinate space~\cite{Voloshin:2003nt}.

The energy-releasing transition of the $\bar{D}^{*0}$ meson to the $\bar{D}^0$ meson is induced by the incoming $D^0$ meson, which involves the inelastic scattering $\bar{D}^{*0} D^0  \to D^0 \bar{D}^0$ through vector exchange, as shown in Fig.~\ref{picture}. To depict the scattering, the following Lagrangians under the heavy quark and chiral symmetries are adopted~\cite{Casalbuoni:1996pg},
\begin{align}\label{eq:lag-p-exch}
  \mathcal{L}_{P^*PV} &=
- i \lambda g_V\varepsilon_{\lambda\alpha\beta\mu}
  (D^{*0\mu\dag}\overleftrightarrow{\partial}^\lambda D^0
  +D^{0\dag}\overleftrightarrow{\partial}^\lambda D^{*0\mu})
  \partial^\alpha{}{V}^\beta\nonumber\\
  &-i\lambda g_V\varepsilon_{\lambda\alpha\beta\mu}
  (\bar{D}^{*0\mu\dag}\overleftrightarrow{\partial}^\lambda
  \bar{D}^0  +\bar{D}^{0\dag}\overleftrightarrow{\partial}^\lambda
  \bar{D}^{*0\mu})\partial^\alpha{}{V}^\beta,\nonumber\\
  	\mathcal{L}_{{PP}\mathbb{V}} &= -i\frac{\beta
	g_V}{2} D^{0\dag}\overleftrightarrow{\partial}^\mu D^0{V}^\mu+i\frac{\beta	g_V}{{2}}\bar{D}^{0\dag}
	\overleftrightarrow{\partial}^\mu \bar{D}^0{V}^\mu,
\end{align}
where $V=\rho^0$ or $\omega$, and the parameters involved here were determined in the literature as $\beta=0.9$, $\lambda=0.56$ GeV$^{-1}$, and $g_V=5.9$~\cite{Casalbuoni:1996pg,Chen:2019asm}. 

Applying standard Feynman rules, the amplitude for the inelastic scattering $\bar{D}^{*0} D^0  \to\bar{D}^0 D^0$ can be written as
\begin{align}
  {\cal A}_{\lambda_{\bar{D}^{*0}}}(\bar{D}^{*0}{D}^0\to \bar{D}^0D^0)
&=2\sqrt{2}\lambda \beta g^2_V \varepsilon_{\lambda\alpha\beta\mu}
p_1^\lambda k^\alpha_1 k^\beta_2\epsilon^{\mu} P(q^2),
\end{align}
with the propagator of exchanged $\rho$ and $\omega$ mesons
\begin{align}
P(q^2)=\sum_{i=\rho, \omega}\frac{1}{q^2-m^2_i}~\frac{m_i^2-\Lambda^2}{q^2-\Lambda^2},
\end{align}
with a standard cutoff $\Lambda=1$~GeV.

With the above amplitudes for  splitting of $X(3872)$ and inelastic scattering  $\bar{D}^{*0}{D}^0\to  \bar{D}^0D^0$, the amplitudes for total reaction $D^0+X\to 2D^0+\bar{D}^0$ can be written as,
\begin{align}
{\cal M}&=\frac{-8\lambda \beta g^2_V\sqrt{m_Xm_{D^*}m_D}}{m_X-m_D+m_{D^*} }P(q^2)
\nonumber\\
&\cdot[\psi({\bm k}_3)\varepsilon_{\lambda\alpha\beta\mu}p_1^\lambda k^\alpha_1 k^\beta_2 \epsilon_X^{\mu}+(2\leftrightarrow 3)].\label{TA}
\end{align}
The $(2\leftrightarrow 3)$ term is an exchange of the momenta for final particle 2 and 3, which is for the second term $D^{*0} \bar{D}^0$ in the wave function in Eq. (2). 

\section{Numerical results}

\begin{figure}[h!] \includegraphics[bb=35 -20 800 580, scale=.33,clip]{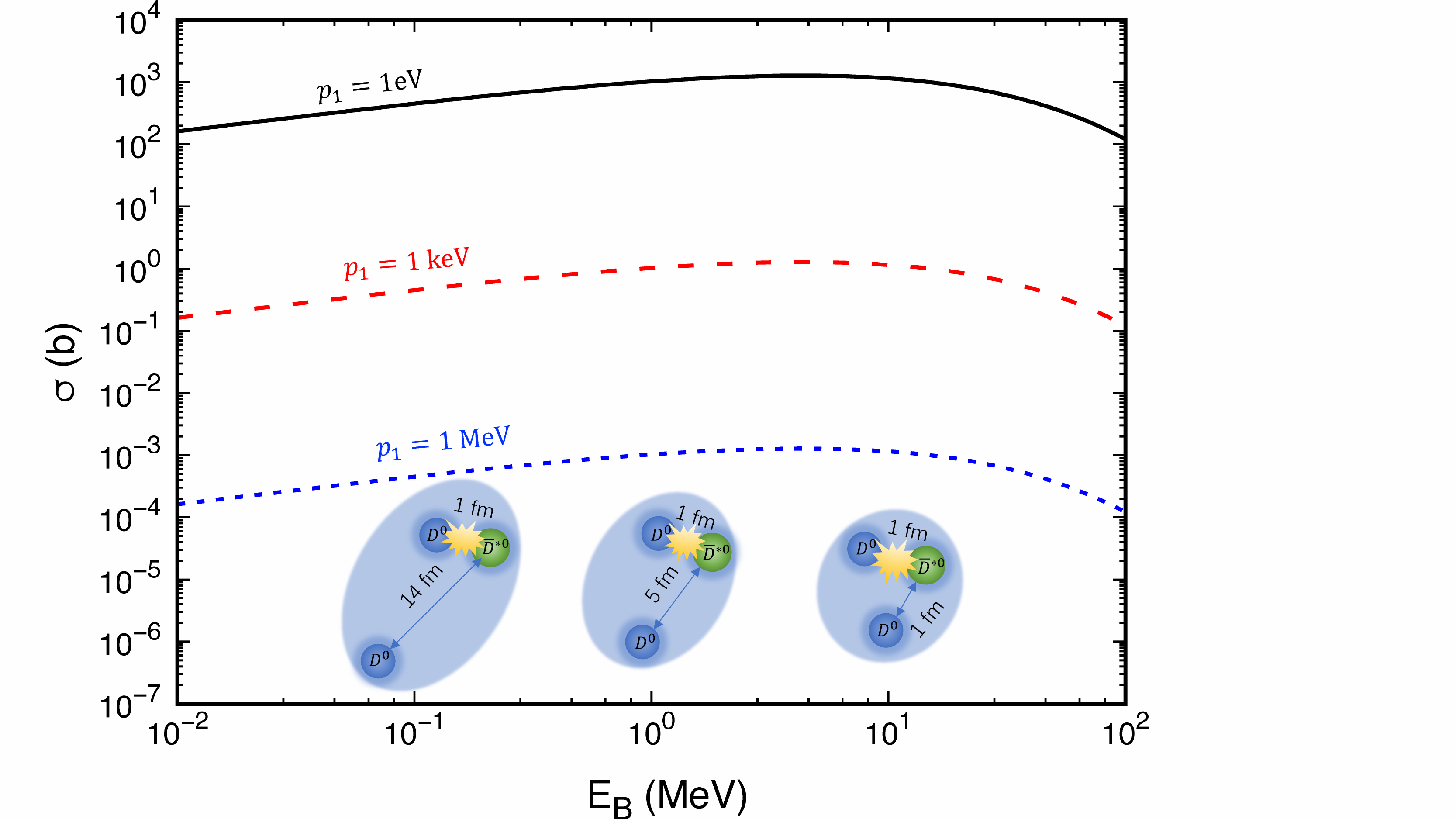}
  \caption{Cross section of the $D^0+X\to 2D^0+\bar{D}^0$ reaction as a function
  of the binding energy. The results for three choices of the momentum of the
  incoming $D^0$ meson $p_1=1$~eV, 1~keV, and 1~MeV, are presented as solid (black), 
  dashed (red), and dotted (blue) curves. The three subfigures show the radii of the molecular
  states with binding energies of 0.1, 1, and 10~MeV. The distance 1~fm between
  collision $D^0$ and $\bar{D}^{*0}$  mesons (two upper mesons) represent the
  interaction range of $\rho/\omega$ exchange.}\label{EB} \end{figure}

The binding energy is an important metric of a hadronic molecular state. Though
there is a very small suggested value of $X(3872)$ listed in the
PDG~\cite{ParticleDataGroup:2020ssz}, we will vary the binding energy from 0.05
to 50~MeV. With such variation, more property of the molecular can be unveiled
and it is also helpful to show the behavior of other molecular states with
different binding energies.  In Fig.~\ref{EB}, for a binding energy of
approximately 0.1~MeV, the cross section is smaller. 
Considering the radius of a
$D$ or $D^*$ meson is smaller than 1~fm, such binding energy means a large
radius, approximately 14~fm, which makes the probability of collision of the
incoming $D^0$ with $D^{*0}$ in $X(3872)$ very small. With increasing binding
energy, the hadronic molecular state becomes more compact, and the cross section
increases. If the binding energy is larger than 10~MeV, which corresponds to a
radius of approximately 1~fm on the same order as the force range of
$\rho/\omega$ exchange, then the cross section decreases again due to the small
size of the molecular state.

\begin{figure}[h!]
\centering
\includegraphics[bb=50 0 750 550, scale=.42,clip]{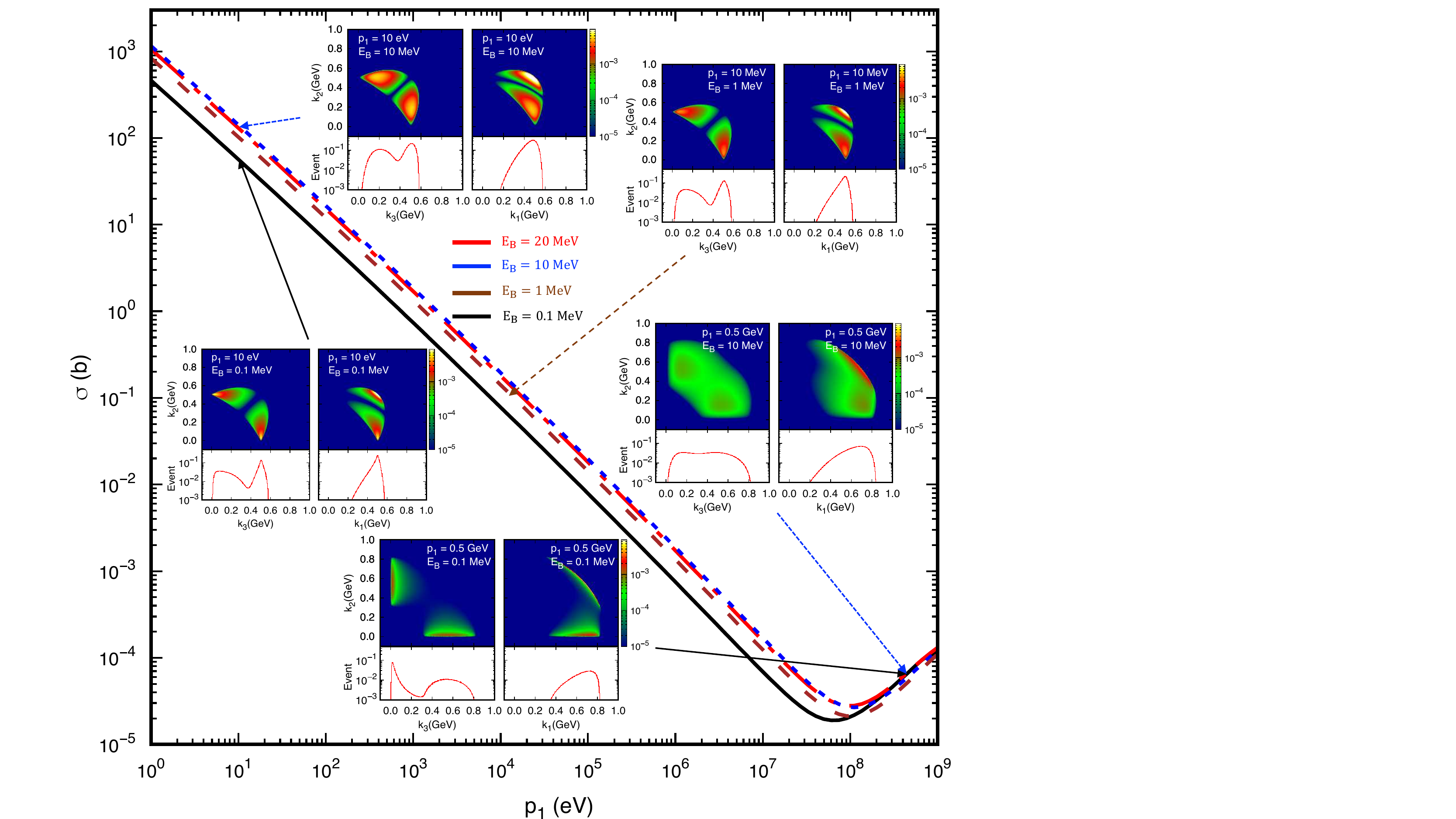}
\caption{Cross section $\sigma$ of the $D^0+X\to 2D^0+\bar{D}^0$ reaction as a
function of the momentum of the incoming $D^0$ meson $p_1$, and the momentum
distributions of final particles. The solid (black), dashed (brown), dotted (blue), and dash-dotted (red) curves are for
the results with four binding energies $E_B=0.1$, 1, 10, and 20~MeV,
respectively. The subfigures show the momentum distributions of final particles.
For each example choice of $\rm p_1$ and $E_B$, the figures represent the $\rm
k_3-k_2$ (left) and $\rm k_1-k_2$ (right) planes, showing the momentum  ${\rm
k}_i=|{\bm k}_i|$  of the final meson $i$. The colorbox means the ratio of event
number in a bin of 0.002 GeV$\times$0.002~GeV to the total number of events. The
two lower panels show the ratio of event number in a bin of 0.002~GeV to the
total number of events against $\rm k_3$ (left) and $\rm k_1$ (right) for
particle 3 and 1 as shown in the Feynman diagram in Fig.~\ref{picture}. The
results here include both component of $X(3827)$ as shown in Eqs.~(\ref{wf}) and
(\ref{TA}). The results are obtained with $10^9$ simulation.  }\label{pb}
\end{figure}

In Fig.~\ref{pb}, the cross section with the variation in the momentum of the
incoming $D^0$ meson $p_1$ and the momentum distribution of final particles are
presented.
The largest cross section is found at a small incoming momentum
$p_1$. With increasing incoming momentum, the cross section decreases almost
linearly and reaches a minimum at approximately 100~MeV. This is reasonable
because the faster incoming $D^0$ meson has a shorter time to interact, which
makes the probability of reaction smaller.

It is important to study the momentum distribution of three final mesons. In
Fig.~\ref{pb}, the distributions separated in to two parts, which corresponds to
two terms of wave function in Eq.~(\ref{wf}) and the amplitudes in
Eq.~(\ref{TA}), and the results for $\rm k_3-k_2$ verifies the symmetry of
particles 2 and 3. With a slow incoming $D^0$ meson, of a momentum of 10~eV for
example, more events with $\rm k_3$ about zero and $\rm k_2$ about 0.5 GeV, or
symmetrically $\rm k_2$ about zero and $\rm k_3$ about 0.5 GeV, can be observed.
It suggests that the $\bar{D}^0$ meson off the struck $\bar{D}^{*0}$ meson
acquires large momentum about 0.5~GeV while the spectator meson, i.e. the
pseudoscalar $D^0$ meson in the $X(3872)$ has a relatively small momentum after
induced fission. If the  incoming energy increases to 0.5~GeV and keeps the
binding energy as 0.1~MeV, the events of the spectator $D^0$ meson concentrates
in small momentum range, exhibited as a sharp peak in the momentum distribution
spectrum of $\rm k_3$. It suggests that with large incoming momentum the
spectator meson in $X(3872)$ is almost unaffected and maintains a very small
momentum as in hadronic molecular state. The $D^0$ meson off the struck
$\bar{D}^{*0}$ meson has a wide distribution around 0.6~GeV. The two terms of
wave function in Eq.~(\ref{wf}) result in the two peaks in the distribution of
the momentum of the $D^0$ or $\bar{D}^0$ meson off the X(3872). The incoming
$D^0$ meson (particle 1 as shown in Feynman diagram) also acquires a larger
momentum $\rm k_1$ about 0.5~GeV with large dispersion.

In addition to $\bar{D}^0$ meson-induced fission-like process, a $\bar{D}^0$ meson can also
induce the fission-like process of $X(3872)$ as follows: \begin{align} &\bar{D}^0+X(3872)\to
\bar{D}^0\bar{D}^{*0}D^0\to {D}^0+2\bar{D}^0.  \end{align} Both $D^0$ and
$\bar{D}^0$ mesons can play the role of neutrons in nuclear fission and
continuously induce chain reactions. Moreover, “cross” chain reactions can also
occur; for example, $\bar{D}^0$ meson-induced fission-like process produces one $D^0$ meson,
which further induces another kind of fission-like process. The fission-like of $X(3872)$ may show
phenomena other than those of nuclear fission, and these phenomena provide
various views and will help us understand the mechanisms of fission-like process more deeply.

\section{Summary}

In this work, we predict an interesting phenomenon, possible fission-like process and chain
reaction  of hadron molecular states. Since only two constituents exist in the
molecular state, the standard type of fission of nuclei is impossible. We notice
that the kinds of hadrons in the molecular states are richer than the nuclei
where only nucleon involved. For example, the $X(3872)$ considered in the
current work, the transition of the $D^*$ meson to $D$ meson make the fission-like process
possible. With the explicit calculation, such reaction exhibit a behavior very
analogous to the nuclear fission especially the rapid increase of the cross section with small incoming momentum. Moreover, the breeding of the $D$ meson of the
reaction, such as $D^0X(3872)\to D^0\bar{D}^0D^0$, makes the chain reaction of a
molecular-state matter possible.  
There can be induced fission-like process with other hadronic molecular states. Here, we list three typical reactions in the strange, charmed, and bottom sectors,
\begin{align}
&K+f_1(1285)\to K\bar{K}^*K\to 2{K}+\bar{K}, \nonumber\\
&D+T_{cc}(3875)\to D{D}^{*}D\to 3{D}, \nonumber\\
&B+Z_b(10610)\to B\bar{B}^*B\to 2{B}+\bar{B}.\nonumber
\end{align}

Due to the short lifetime of the molecular states, such reactions should not be
observed directly in near future. However, it may exhibit its effect in
some scenes. Taking $X(3872)$ as an example, Quark-Gluon-Plasma will be produced
through nuclear collisions in which there will be a large number of charm
mesons~\cite{Zhang:2020dwn}. Charmed mesons can form many $X(3872)$ particles through strong
interaction, and then further react with the remaining charmed $D$ mesons to
make the proposed fission-like process of hadron molecule happen. 

\acknowledgments

This work is supported by the China National Funds for Distinguished Young Scientists under Grant No. 11825503, the National Key Research and Development Program of China under Contract No. 2020YFA0406400, the 111 Project under Grant No. B20063, the National Natural Science Foundation of China under Grant No. 12047501, No. 12175091, No. 11965016, No. 11775050 and No. 11775050, CAS Interdisciplinary Innovation Team, and the Fundamental Research Funds for the Central Universities under Grants No. lzujbky-2021-sp24.


\begin{thebibliography}{90}

%\cite{Belle:2003nnu}
\bibitem{Belle:2003nnu}
S.~K.~Choi \textit{et al.} [Belle],
``Observation of a narrow charmonium - like state in exclusive $B^\pm\to K^\pm \pi^+ \pi^- J /\psi$ decays,''
Phys. Rev. Lett. \textbf{91}, 262001 (2003)
%doi:10.1103/PhysRevLett.91.262001
%[arXiv:hep-ex/0309032 [hep-ex]].
%2000 citations counted in INSPIRE as of 15 Sep 2021

%\cite{Tornqvist:2004qy}
\bibitem{Tornqvist:2004qy}
N.~A.~Tornqvist,
``Isospin breaking of the narrow charmonium state of Belle at 3872-MeV as a deuson,''
Phys. Lett. B \textbf{590}, 209-215 (2004)
%doi:10.1016/j.physletb.2004.03.077
%[arXiv:hep-ph/0402237 [hep-ph]].
%542 citations counted in INSPIRE as of 15 Sep 2021




%\cite{Liu:2008fh}
\bibitem{Liu:2008fh}
Y.~R.~Liu, X.~Liu, W.~Z.~Deng and S.~L.~Zhu,
``Is $X(3872) $ Really a Molecular State?,''
Eur. Phys. J. C \textbf{56}, 63-73 (2008)
%doi:10.1140/epjc/s10052-008-0640-4
%[arXiv:0801.3540 [hep-ph]].
%150 citations counted in INSPIRE as of 15 Sep 2021


%\cite{Meitner:1939gwm}
\bibitem{Meitner:1939gwm}
L.~Meitner and O.~R.~Frisch,
``Disintegration of Uranium by Neutrons: a New Type of Nuclear Reaction,''
Nature \textbf{143}, no.3615, 239-240 (1939)
%doi:10.1038/143239a0
%172 citations counted in INSPIRE as of 15 Sep 2021

%\cite{Chen:2020aos}
\bibitem{Chen:2020aos}
H.~X.~Chen, W.~Chen, R.~R.~Dong and N.~Su,
``$X_0$(2900) and $X_1$(2900): Hadronic Molecules or Compact Tetraquarks,''
Chin. Phys. Lett. \textbf{37}, no.10, 101201 (2020)
%doi:10.1088/0256-307X/37/10/101201
%[arXiv:2008.07516 [hep-ph]].
%46 citations counted in INSPIRE as of 25 Jul 2022

%\cite{Liu:2021pdu}
\bibitem{Liu:2021pdu}
M.~Z.~Liu and L.~S.~Geng,
``Prediction of an \ensuremath{\Omega}bbb\ensuremath{\Omega}bbb Dibaryon in the Extended One-Boson Exchange Model,''
Chin. Phys. Lett. \textbf{38}, no.10, 101201 (2021)
%doi:10.1088/0256-307X/38/10/101201
%[arXiv:2107.04957 [hep-ph]].
%6 citations counted in INSPIRE as of 25 Jul 2022

%\cite{Li:2021zbw}
\bibitem{Li:2021zbw}
N.~Li, Z.~F.~Sun, X.~Liu and S.~L.~Zhu,
``Perfect DD* Molecular Prediction Matching the Tcc Observation at LHCb,''
Chin. Phys. Lett. \textbf{38}, no.9, 092001 (2021)
%doi:10.1088/0256-307X/38/9/092001
%[arXiv:2107.13748 [hep-ph]].
%28 citations counted in INSPIRE as of 25 Jul 2022

%\cite{He:2022rta}
\bibitem{He:2022rta}
J.~He and X.~Liu,
``The quasi-fission phenomenon of double charm $T_{cc}^+$ induced by nucleon,''
Eur. Phys. J. C \textbf{82}, no.4, 387 (2022)
%doi:10.1140/epjc/s10052-022-10363-4
%[arXiv:2202.07248 [hep-ph]].
%4 citations counted in INSPIRE as of 03 May 2022



%\cite{Braaten:2004fk}
\bibitem{Braaten:2004fk}
E.~Braaten, M.~Kusunoki and S.~Nussinov,
``Production of the X(3870) in B meson decay by the coalescence of charm mesons,''
Phys. Rev. Lett. \textbf{93}, 162001 (2004)
%doi:10.1103/PhysRevLett.93.162001
%[arXiv:hep-ph/0404161 [hep-ph]].
%69 citations counted in INSPIRE as of 29 Jul 2021

%\cite{Weinberg:1962hj}
\bibitem{Weinberg:1962hj}
S.~Weinberg,
``Elementary particle theory of composite particles,''
Phys. Rev. \textbf{130}, 776-783 (1963)
%doi:10.1103/PhysRev.130.776
%540 citations counted in INSPIRE as of 15 Sep 2021

%\cite{He:2011ed}
\bibitem{He:2011ed}
J.~He and X.~Liu,
``The open-charm radiative and pionic decays of molecular charmonium Y(4274),''
Eur. Phys. J. C \textbf{72}, 1986 (2012)
%doi:10.1140/epjc/s10052-012-1986-1
%[arXiv:1102.1127 [hep-ph]].
%43 citations counted in INSPIRE as of 08 Aug 2021

  %\cite{Voloshin:2003nt}
\bibitem{Voloshin:2003nt}
M.~B.~Voloshin,
``Interference and binding effects in decays of possible molecular component of X(3872),''
Phys. Lett. B \textbf{579}, 316-320 (2004)
%doi:10.1016/j.physletb.2003.11.014
%[arXiv:hep-ph/0309307 [hep-ph]].
%290 citations counted in INSPIRE as of 08 Aug 2021



%\cite{Casalbuoni:1996pg}
\bibitem{Casalbuoni:1996pg}
R.~Casalbuoni, A.~Deandrea, N.~Di Bartolomeo, R.~Gatto, F.~Feruglio and G.~Nardulli,
``Phenomenology of heavy meson chiral Lagrangians,''
Phys. Rept. \textbf{281}, 145-238 (1997)
%doi:10.1016/S0370-1573(96)00027-0
%[arXiv:hep-ph/9605342 [hep-ph]].
%547 citations counted in INSPIRE as of 29 Sep 2020


%\cite{Chen:2019asm}
\bibitem{Chen:2019asm}
R.~Chen, Z.~F.~Sun, X.~Liu and S.~L.~Zhu,
``Strong LHCb evidence supporting the existence of the hidden-charm molecular pentaquarks,''
Phys. Rev. D \textbf{100}, no.1, 011502 (2019)
%doi:10.1103/PhysRevD.100.011502
%[arXiv:1903.11013 [hep-ph]].
%79 citations counted in INSPIRE as of 29 Sep 2020
  
%\cite{ParticleDataGroup:2020ssz}
\bibitem{ParticleDataGroup:2020ssz}
P.~A.~Zyla \textit{et al.} [Particle Data Group],
``Review of Particle Physics,''
PTEP \textbf{2020}, no.8, 083C01 (2020)
%doi:10.1093/ptep/ptaa104
%1938 citations counted in INSPIRE as of 15 Sep 2021

%\cite{Zhang:2020dwn}
\bibitem{Zhang:2020dwn}
H.~Zhang, J.~Liao, E.~Wang, Q.~Wang and H.~Xing,
``Deciphering the Nature of X(3872) in Heavy Ion Collisions,''
Phys. Rev. Lett. \textbf{126}, no.1, 012301 (2021)
%doi:10.1103/PhysRevLett.126.012301
%[arXiv:2004.00024 [hep-ph]].
%18 citations counted in INSPIRE as of 25 Jul 2022

\end{thebibliography}
\end{document}